# Extremely large magnetoresistance in the hourglass Dirac loop chain metal β-ReO$_2$

Daigorou Hirai[1]*, Takahito Anbai[1], Shinya Uji[2], Tamio Oguchi[3,4], and Zenji Hiroi[1]

[1]*Institute for Solid State Physics, University of Tokyo, Kashiwa, Chiba 277-8581, Japan*
[2]*National Institute for Materials Science, Tsukuba, Ibaraki 305-0003, Japan*
[3]*Institute of Scientific and Industrial Research, Osaka University, Ibaraki, Osaka 567-0047, Japan*
[4]*Center for Spintronics Research Network, Osaka University, Toyonaka, Osaka 560-8531, Japan*

The transport and thermodynamic properties of β-ReO$_2$ crystallizing in a nonsymmorphic structure were studied using high-quality single crystals. An extremely large magnetoresistance (XMR) reaching 22,000% in a transverse magnetic field of 10 T at 2 K was observed. However, distinguished from other topological semimetals with low carrier densities that show XMR, β-ReO$_2$ has a high electron carrier density of $1 \times 10^{22}$ cm$^{-3}$ as determined by Hall measurements and large Fermi surfaces in the electronic structure. In addition, a small Fermi surface with a small effective mass was evidenced by de Haas–van Alphen oscillation measurements. The previous band structure calculations [S. S. Wang, *et al*., Nat. Commun. **8**, 1844 (2017)] showed that two kinds of loops made of Dirac points of hourglass-shaped dispersions exist and are connected to each other by a point to form a string of alternating loops, called the Dirac loop chain (DLC), which are protected by the multiple glide symmetries. Our first-principles calculations revealed the complex Fermi surfaces with the smallest one corresponding to the observed small Fermi surface, which is just located near the DLC. The XMR of β-ReO$_2$ is attributed to the small Fermi surface and thus is likely caused by the DLC.

## 1. Introduction

A wide variety of physical properties appear in 5$d$ transition metal compounds owing to competition and cooperation between the spin–orbit interactions (SOIs) as large as 0.4 eV, crystal fields, and electron correlations.[1] Recent extensive materials investigations revealed the novel electronic properties of 5$d$ compounds.[2,3] Particularly observed in rhenium compounds are superconductivity in Cd$_2$Re$_2$O$_7$ and Hg$_x$ReO$_3$,[4–7] multipole orders in Cd$_2$Re$_2$O$_7$ and Ba$_2$MgReO$_6$,[8–12] half-metallic states above room temperature in $A_2B$ReO$_6$,[13,14] pleochroism in Ca$_3$ReO$_5$Cl$_2$,[15] and frustrated magnetism in $A_3$ReO$_5$Cl$_2$.[16–18]

For *sp* electron systems, on the other hand, unconventional properties derived from the topology of Fermi surface have attracted considerable attention. The topological semimetal has a singular point at a crossing of two bands with linear dispersion in the band structure, which is protected by symmorphic crystal symmetry such as a mirror plane. When such a band crossing exists near the Fermi energy ($E_F$), the quasiparticles behave like relativistic Dirac or Weyl fermions and cause exotic transport properties such as ultra-high mobility and extremely large magnetoresistance (XMR).[19–22] For example, the Dirac semimetal Cd$_3$As$_2$ showed mobility as high as $9 \times 10^6$ cm$^2$ V$^{-1}$ s$^{-1}$ at 5 K and XMR of $1.3 \times 10^5$% at 9 T.[21] Such anomalous properties may not appear in 5$d$ electron systems with strong SOIs, because in most cases, a band crossing is unstable against SOIs and thus a finite gap opens to remove the topological nature.

Recently, a specific type of band crossing protected by nonsymmorphic symmetry such as a glide plane has been focused because it is robust against SOIs.[23–26] Let us consider two high-symmetry $k$-points, $k_1$ and $k_2$, that are protected by a glide symmetry as well as time-reversal symmetry (Fig. 1). Any band at $k_1$ or $k_2$ is fourfold degenerate owing to the two symmetry requirements, while it splits into a pair of spin-degenerate branches of different irreducible representations on a line between $k_1$ and $k_2$. Thus, it can happen that two quartets A and B at $k_1$ exchange their spin-degenerate branches at $k_2$ to recover the quartet degeneracy. This exchange should generate a crossing point with fourfold degeneracy with linear dispersion, that is, the Dirac point. As a result, such an hourglass band dispersion as illustrated in Fig. 1 is obtained. Note that the Dirac point at the hourglass-type neck crossing point is protected by glide symmetry and is not split by SOIs. Hence, it is a good ingredient to introduce topology-related phenomena into $5d$ compounds. A more interesting case is predicted when multiple nonsymmorphic symmetries exist in a crystal structure. In such a case, Dirac points can be aligned to form a loop and moreover a Dirac loop chain (DLC) made of multiple Dirac loops, as schematically depicted in Fig. 1. Experimentally, an hourglass-type dispersion was observed on the surface of KHg$X$ ($X$ = As, Sb, Bi).[27] However, little is known about hourglass-type dispersions protected by nonsymmorphic symmetry in bulk crystals and related topological phenomena. Very recent electronic structure calculations showed that β-ReO$_2$ has such a DLC protected by the glide symmetry.[28]

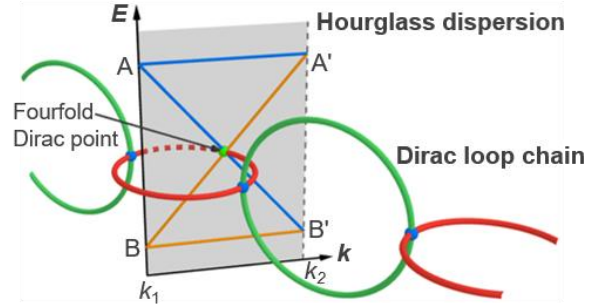

**Fig. 1.** (Color online) Schematic representation of the hourglass dispersion and Dirac loop chain. Glide symmetry in a crystal structure can produce an hourglass-shaped band dispersion with a fourfold degenerate Dirac point at the neck crossing point between two high symmetry $k$ points, $k_1$ and $k_2$. In β-ReO$_2$ having three glide planes in its crystal structure, Dirac points trace out two kinds of Dirac loops (red and green rings) which are connected to each other by a point to form a string of alternating loops, that is, the Dirac loop chain.

Three polymorphs of ReO$_2$ are known: rutile-type, α-phase, and β-phase. The rutile form was obtained only when synthesized by metathesis reactions or in the form of thin film under epitaxial strain.[29,30] The α-phase was synthesized below 600 K, and the β-phase was obtained at higher synthetic temperatures.[31] The α-phase crystallizes in a MoO$_2$-type distorted rutile structure, and the β-phase takes an orthorhombic α-PbO$_2$ type structure with the space group $Pbcn$ (#60).[32,33] In all the three polymorphs of ReO$_2$, the rhenium ion is octahedrally coordinated by six oxide atoms, while the octahedra are connected in different ways. In β-ReO$_2$ as depicted in Fig. 2a, each octahedron shares edges with two neighboring octahedra to form a zigzag chain extending in the $c$ direction. Each chain shares vertices with four octahedra in the nearby zigzag chains, resulting in a three-dimensional crystal structure. Note that there are three glide planes in the crystal structure: a $b$-glide plane perpendicular to the $a$ axis, a $c$-glide plane perpendicular to the $b$ axis, and an $n$-glide plane perpendicular to the $c$ axis.

According to the band structure calculations taking SOIs into account,[28] β-ReO$_2$ is a metal with large Fermi surfaces originating from the $5d$ states of rhenium strongly hybridized with the $2p$ states of oxygen. There exist two kinds of Dirac loops surrounding high symmetry $k$ points, which are made of neck-crossing points of the hourglass dispersion protected by the glide planes. Moreover, it is predicted from symmetry considerations that the Dirac loops form a DLC running along the $k_y$ direction, which is the consequence of symmetry protection by the two orthogonal $c$- and $n$-glide planes, although the actual shape of the DLC has not been clarified. Importantly, this DLC exists in the vicinity of $E_F$[28] and thus is expected to be responsible for some physical properties. Therefore, β-ReO$_2$ is a promising candidate for a new topological phase called the hourglass DLC metal. In contrast to $sp$-electron based topological materials, heavy $5d$ electrons must play a dominant role in the physical properties of β-ReO$_2$, which may lead to novel electronic properties.

In the previous experimental study, β-ReO$_2$ showed a metallic conductivity without signatures for superconductivity nor a magnetic transition down to 4.2 K.[31] No physical properties suggesting the existence of DLC were reported. However, we point out that the quality of crystals used in the previous study was poor, as evidenced by the small residual-resistivity-ratio (RRR) of 8. In this study, we succeeded in growing high-quality single crystals with RRR ~ 1000 and performed detailed transport and thermodynamic measurements. As a result, XMR as large as 22,000% was observed at 10 T and 2 K. Hall measurements and band structure calculations revealed large Fermi surfaces with high carrier density, and quantum oscillations in magnetization measurements revealed a small Fermi pocket with a small effective mass. Interestingly, this Fermi pocket is located close to the DLC determined by our first-principles calculations. Thus, it is likely that the quasiparticles originating from the DLC is responsible for the observed XMR. β-ReO$_2$ may be a promising hourglass Dirac chain metal candidate.

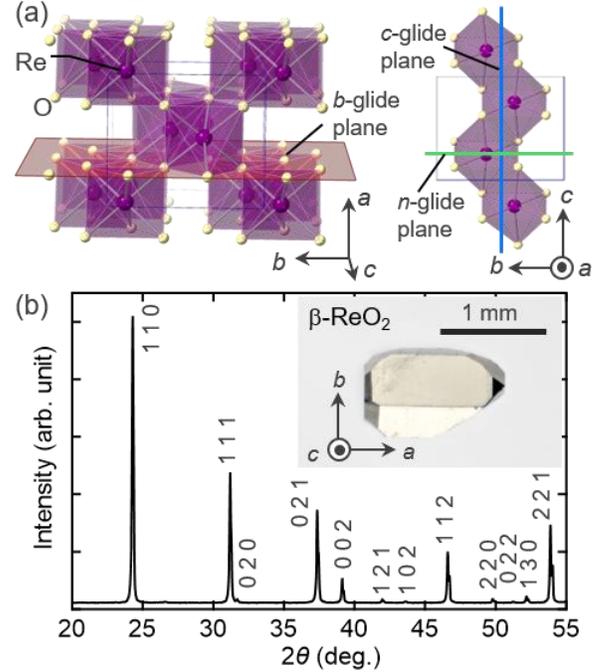

**Fig. 2.** (Color online) (a) Crystal structure of β-ReO$_2$. A string of ReO$_6$ octahedra containing a zigzag chain of Re atoms along the $c$ axis is shown on the right. There are three kinds of glide planes in the crystal structure: $b$-, $c$-, and $n$-glide planes perpendicular to the $a$, $b$, and $c$ axes, respectively. (b) Powder XRD pattern of a crushed crystal of β-ReO$_2$. The diffraction indices are based on the α-PbO$_2$-type structure. The inset shows a photograph of a typical single crystal of β-ReO$_2$.

## 2. Experimental

Single crystals of β-ReO$_2$ were grown by the chemical vapor transport method given in the literature.[34] Firstly, Re$_2$O$_7$ and Re metal were mixed in a molar ratio of 1:3 in an argon-filled glove box. A pellet made from the powder was sealed in a quartz tube and sintered at 1073 K for 24 hours. The sintered sample was confirmed to be β-ReO$_2$ with the α-PbO$_2$-type structure by X-ray diffraction (XRD) measurements using Cu-Kα radiation (Rigaku RINT-2000), which is consistent with the previous report that β-phase was obtained when sintered above 600 K.[31] The β-ReO$_2$ powder was sealed in a quartz tube with iodine as a transport agent and placed in a furnace with a temperature gradient of 1023–1123 K for 20 days. As-grown crystals were covered with small red crystals of ReO$_3$, which were removed by ultrasonic cleaning in acetone. The thus-obtained silver crystals had a rod-shape with the maximum size of 2 × 0.8 × 0.8 mm$^3$ (Fig. 2b). They were stable in air.

The obtained single crystals were crushed and examined by powder XRD measurements, which confirmed the β-ReO$_2$-type structure without impurities. As shown in Fig. 2b, the powder XRD pattern is in good agreement with the reported pattern of β-ReO$_2$. The refined lattice constants of $a$ = 4.8082(9) Å, $b$ = 5.6382(12) Å, and $c$ = 4.6005(11) Å are almost the same as the reported values of $a$ = 4.8094(5) Å, $b$ = 5.6433(5) Å, and $c$ = 4.6007(5) Å.[33] The orientation of a single crystal was determined by single-crystal XRD measurements using Mo-Kα radiation on a diffractometer (Rigaku, R-AXIS RAPID IP). The direction of the long axis of the rod-shaped crystal corresponds to the $a$ axis, and the $b$, $c$, and (011) planes appear as crystal habits on the side surfaces (Fig. 2b). In some of the samples, twins with boundary planes parallel to the $a$ axis were formed.

Electrical resistivity, Hall, and heat capacity measurements were performed in a physical properties measurement system (PPMS, Quantum Design). Electrical resistivity was measured by the four-probe method, and the Hall effect was measured by the five-probe method. The electrical contacts for the transport measurements were made by silver paste (DuPont 4922N, DuPont Inc.). For both resistivity and Hall measurements, an electrical current was flown along the $a$ direction and magnetic fields between –10 and 10 T were applied in the $c$ direction. Hall measurements were performed at constant temperatures with sweeping magnetic field. $\rho_{yx}(B)$ was obtained as $\rho_{yx}(B) = [\rho_{yx}(B) - \rho_{yx}(-B)]/2$, which eliminated a $\rho_{xx}(B)$ contribution due to a misalignment of the contacts. Heat capacity was measured by the relaxation method using one single crystal of 23.3 mg weight. Good thermal contact between the crystal and holder was made using grease (Apiezon-N, M & I Materials Ltd). Magnetic susceptibility measurements were carried out on a single crystal in a magnetic properties measurement system (MPMS3, Quantum Design). The crystal was mounted on a quartz rod with varnish (GE7031, General Electric Company), and a magnetic field was applied in the direction parallel and perpendicular to the $a$ axis.

First-principles electronic structure calculations of β-ReO$_2$ were performed by using the all-electron full-potential linearized augmented plane wave (FLAPW) method[35–37] implemented in the HiLAPW code[38] on the basis of the generalized gradient approximation to the density-functional theory (DFT).[39] The SOIs were self-consistently taken into account for the valence and core states by the second variation scheme.[40] The energy cutoffs of 20 and 160 Ry were used for wavefunction and potential expansions, respectively. The lattice constants and atomic coordinates were fetched from the Materials Project site[41] and were confirmed to be reliable by our FLAPW calculations within less than 1% in lattice constants and 5 mRy/Bohr in the atomic forces. Brillouin-zone sampling was made with Γ-centered 16×16×16 (32×32×32) mesh points in the self-consistent (state density) calculations.

## 3. Results

### 3.1 Transport properties

Resistivity measurements revealed that high-quality single crystals were obtained, as shown in Fig. 3. $\rho_{xx}(T)$ in zero magnetic field shows metallic behavior with a pronounced decrease below 100 K. The RRR, defined as $\rho_{xx}(300\,K)/\rho_{xx}(2\,K)$, is 990, much larger than the previously reported RRR of 8;[31] there were scatters in the value of RRR among single crystals from the same preparation badge, and the average was about 400. Furthermore, the residual resistivity $\rho_{xx}(2\,K)$ is very small, 206 nΩ cm, which is comparable to $\rho_{xx}(4.2\,K)$ = 291 nΩ cm for a Bi single crystal grown from a 99.9999% high-purity raw material.[42] The temperature dependence of $\rho_{xx}(T)$ below 20 K is well reproduced by the form of $\rho_{xx}(T) = \rho_0 + AT^2$ with $A = 1.37\times10^{-3}$ μΩ cm K$^{-2}$ (Fig. 3a). This $T^2$ dependence suggests that a strong scattering mechanism such as electron–electron scattering is dominant, rather than the usual scattering mechanism by long-wavelength phonons. A superconducting transition was not observed above 1.8 K.

When a transverse magnetic field is applied perpendicular to the current direction ($B \parallel [001]$, $I \parallel [100]$), the resistivity increases significantly below 100 K, resulting in dips at 20 K (1 T) and 43 K (10 T); the dip temperature increases almost proportionally to $B$. As shown in Fig. 3b, the isothermal magnetoresistance (MR), defined as $[R(B) - R(0)]/R(0)$, at 2 K increases rapidly with increasing magnetic field, and eventually reaches a large positive value of 22,000% at 10 T. The MR initially increases slowly at low fields and then rapidly at high fields. Fits to the power law yield magnetic field dependences of $B^{1.3}$ at low fields below 1 T and $B^{1.7}$ between 4 and 10 T (Fig. 3b, inset). Note that a $B$-linear MR was theoretically predicted for Dirac fermion systems,[43] while MR ∝ $B^2$ is expected for compensated semimetals.[44]

Hall measurements were performed to determine the type and density of carriers. Figure 4 shows Hall resistivities $\rho_{yx}(B)$ which are almost linear against $B$ with negative slopes at all temperatures, indicating the dominance of electron carriers. A high carrier density of $1\times10^{22}$ cm$^{-3}$ is estimated from the Hall coefficient $R_H(T)$ assuming a single-carrier model as $n(T) = 1/[eR_H(T)]$. This value is comparable to those of simple metals with large Fermi surfaces; for example, $8.9\times10^{22}$ cm$^{-3}$ for copper.[45] Moreover, the temperature dependence of $n(T)$ is small, as shown in the inset of Fig. 4. Both the $B$-linear $R_H(T)$ and the temperature-independent $n(T)$ suggest that conduction electrons originating from large Fermi surfaces dominate the Hall response in β-ReO$_2$.

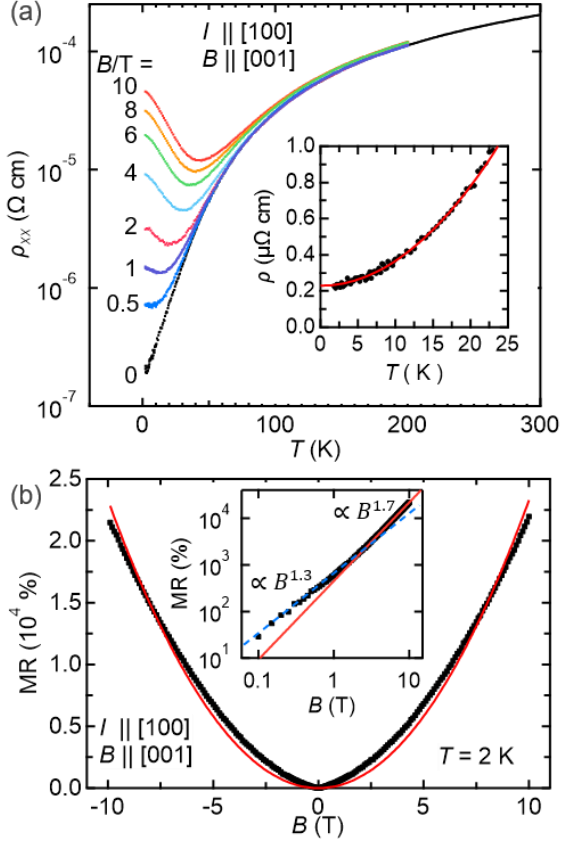

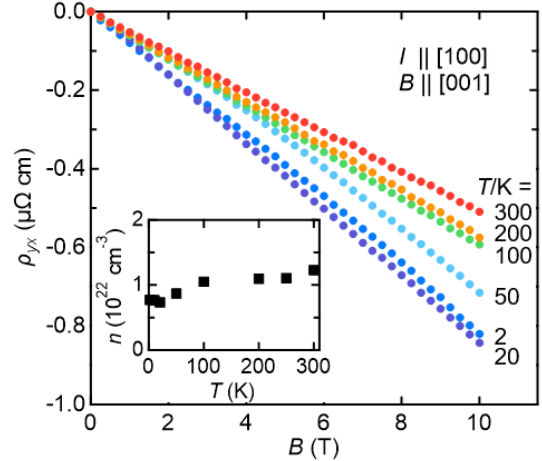

**Fig. 4.** (Color online) Hall resistivity $\rho_{yx}(B)$ of β-ReO$_2$ measured at various temperatures from 2 to 300 K. The inset shows the temperature dependence of carrier density $n(T)$.

**Fig. 3.** (Color online) (a) Temperature dependences of resistivity $\rho_{xx}$ measured with the electrical current $I$ running along the [100] direction at various magnetic fields along the [001] direction. The inset shows a linear-scale plot of the zero-field data at the low-temperature range in which the red solid line is a fit to the equation $\rho_{xx}(T) = \rho_0 + AT^2$. (b) Magnetic field dependence of magnetoresistance (MR) measured at 2 K. The red solid line is a parabolic curve (MR $\propto B^2$) for comparison. The MR is also plotted on the log log scale in the inset. The blue broken and red solid lines are fits to the power law with the exponents of 1.3 and 1.7, respectively.

*3.2 Fundamental electronic states*

Heat capacity and magnetic susceptibility measurements revealed the fundamental electronic states of β-ReO$_2$. The inset of Fig. 5 shows a $C/T$ versus $T^2$ plot below 10 K. A fit to the form of $C(T) = \gamma_{exp}T + \beta T^3$ yields $\gamma_{exp}$ = 3.178(5) mJ K$^{-2}$ mol$^{-1}$ and $\beta$ = 0.03326(9) mJ K$^{-4}$ mol$^{-1}$. The Sommerfeld coefficient $\gamma_{exp}$ is slightly smaller than 6.45 mJ K$^{-2}$ mol$^{-1}$ for ReO$_3$[46] and 4.9 mJ K$^{-2}$ mol$^{-1}$ for Hg$_x$ReO$_3$.[7] There is a small enhancement by 8.5% compared with the calculated value from the density of states (DOS) at $E_F$ in our first-principles calculations, $\gamma_{calc}$ = 2.93 mJ K$^{-2}$ mol$^{-1}$. The Debye temperature $\Theta_D$ given by $\Theta_D = (12\pi^4 NR/\beta)^{1/3}$ is calculated to be 560 K, where $N$ is the number of atoms per formula unit and $R$ is the gas constant. This value is close to those of related rhenium oxides: 460 K for ReO$_3$[46] and 458 K for Cd$_2$Re$_2$O$_7$.[4] No anomalies corresponding to phase transitions were observed in heat capacity between 2 and 300 K.

The magnetic susceptibility of a single crystal of β-ReO$_2$ is positive with little temperature dependence and nearly isotropic with a slight difference above 100 K for magnetic fields parallel and perpendicular to the $a$ axis, which is ascribed to the Pauli paramagnetism of conduction electrons. By subtracting the core diamagnetic responses of $-2.8\times10^{-5}$ cm$^3$ mol$^{-1}$ for Re$^{4+}$ and $-2.4\times10^{-5}$ cm$^3$ mol$^{-1}$ for O$^{2-}$ in the literature[47] from $\chi$ = 4.7×10$^{-5}$ cm$^3$ mol$^{-1}$ at 100 K, the Pauli paramagnetic susceptibility is estimated to be $\chi_{exp}$ = 9.9×10$^{-5}$ cm$^3$ mol$^{-1}$. This value is 2.5 times larger than $\chi_{calc}$ = 4.0×10$^{-5}$ cm$^3$ mol$^{-1}$ from the band structure calculations. However, the actual enhancement may be smaller because the magnetic susceptibility must contain an experimental orbital paramagnetic component; in fact, orbital paramagnetism as large as 1.6×10$^{-4}$ cm$^3$ mol-Re$^{-1}$ was observed for Cd$_2$Re$_2$O$_7$ (Re$^{5+}$: $5d^2$).[48] It is noted that there is an increase in the magnetic susceptibility at low temperatures below 50 K, the origin of which is not clear. Provided that it comes from localized impurity spins, a Curie fit gives a Curie constant of 1.99(1)×10$^{-4}$ cm$^3$ K mol$^{-1}$, which corresponds to only 0.053% spin-1/2. It is also noted that the $\chi$ for $B \parallel$ [100] exhibits a peak at 5 K, which is not due to a long-range magnetic order but due to quantum oscillations, as will be mentioned later.

Using the obtained Sommerfeld coefficient and the Pauli paramagnetic susceptibility, the Wilson ratio $R_W$ = $(\pi^2 k_B^2)/(3\mu_B^2)(\chi/\gamma)$ is calculated to be 2.26, which exceeds 2 expected for strongly correlated electron metals. However, the actual $R_W$ may be smaller because of the not-considered contribution of orbital paramagnetism. On the other hand, the observed $T^2$ resistivity seems to evidence electron correlations. The Kadowaki–Woods ratio, $A/\gamma^2$, is evaluated to be 136 μΩ cm K$^2$ J$^{-2}$ mol$^2$. This value is larger than the typical value of 10 μΩ cm K$^2$ J$^{-2}$ mol$^2$ for heavy-fermion compounds[49] and is comparable to those for strongly correlated oxides such as La$_{1-x}$Sr$_x$CuO$_4$,[50] LiV$_2$O$_4$,[51] and Na$_x$CoO$_2$.[52] One expects a sizable electron correlation effect for β-ReO$_2$ because the $t_{2g}$ manifold is half-filled, as in the pyrochlore oxide Cd$_2$Os$_2$O$_7$ (Os$^{5+}$: 5$d^3$) which exhibits a metal–insulator transition owing to large electron correlations.[53]

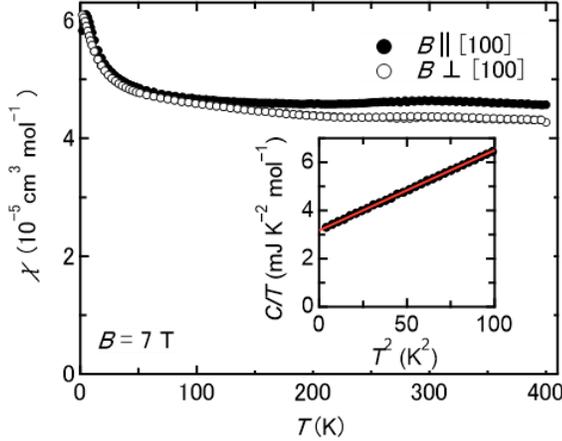

**Fig. 5.** (Color online) Temperature dependences of magnetic susceptibilities for β-ReO$_2$ with applied fields of 7 T parallel (filled circles) and perpendicular (open circles) to the $a$ axis. Inset shows heat capacity divided by temperature ($C/T$) as a function of $T^2$ below 10 K.

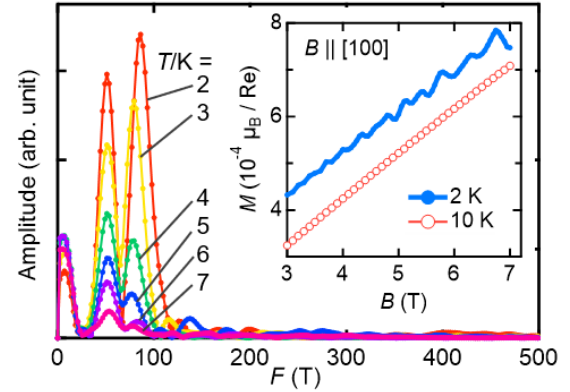

**Fig. 6.** (Color online) Fast Fourier transformation of the oscillations in magnetization in the range of magnetic fields of 3–7 T at various temperatures between 2 and 7 K for β-ReO$_2$. The field dependences of magnetization at 2 and 10 K are shown in the inset.

### 3.3 Quantum oscillations in magnetization

Quantum oscillations were observed in magnetization measurements at relatively high temperatures and low magnetic fields. As shown in Fig. 6, the $MH$ curve is linear to $B$ at 10 K, while it shows an oscillation at 2 K due to the de Haas–van Alphen (dHvA) effect. After subtracting a linear background determined at $B = 3$–7 T, a fast Fourier transform of the data against $1/B$ revealed two major oscillatory components at $F = 51$ and 79 T. These correspond to extremal cross sections of Fermi surface of $A_F = 0.0048$ and 0.0076 Å$^{-2}$, respectively, using Onsager's relation $F = (\Phi_0/2\pi^2)A_F$, where $\Phi_0$ is the magnetic quantum flux. Note that they correspond to very small areas of 1.3 and 2.0% of the first Brillouin zone, respectively.

The dHvA oscillations were reduced with increasing temperature and disappeared around 8 K, as shown in Fig. 6. The temperature dependences of the amplitudes are well fitted by the Lifshitz–Kosevich formula, in which the amplitude $Amp$ is proportional to the thermal damping factor $R_T$ and the Dingle damping factor $R_D$, as $Amp \propto R_T R_D$; $R_T = (2\pi^2 k_B T m^*/eB\hbar)/\sinh(2\pi^2 k_B T m^*/eB\hbar)$ and $R_D = \exp(-2\pi^2 k_B T_D m^*/eB\hbar)$, where $k_B$ is the Boltzmann constant, $m^*$ is the effective mass, and $T_D$ is the Dingle temperature.[54]

The obtained effective masses are 0.23$m_0$ and 0.40$m_0$ for $F = 51$ and 79 T, respectively, where $m_0$ is the bare mass of an electron. Thus, the light quasiparticles originating from small Fermi pockets are the sources of the observed dHvA oscillations. These quasiparticles are heavier than those in topological semimetals showing XMR, such as 0.044$m_0$ in Cd$_3$As$_2$[20] and 0.076 $m_0$ in NbP,[22] while comparable to 0.11 ~ 0.27$m_0$ in ZrSiS,[55] and lighter than 0.64 ~ 0.68$m_0$ in PtBi$_2$.[56] Further quantum oscillation experiments using the magnetic torque at lower temperatures and higher magnetic fields will be reported elsewhere.

### 3.4 Electronic structure calculations

First-principles calculations were carried out to understand the electronic and physical properties of β-ReO$_2$. Our calculations consistently reproduced the previously-reported results[28] and revealed the actual shape of DLC in the band structure. The dominant electronic states near $E_F$ originate from the half-filled $t_{2g}$ manifolds of 5$d$ orbitals of the Re$^{4+}$ ions, which are strongly hybridized with the oxygen 2$p$ orbitals. The partial DOS plot in the inset of Fig. 7 shows that the $d_{xy}$ orbital-derived band is just located at around $E_F$, while the $d_{yz}$- and $d_{zx}$-derived bands split above and below $E_F$. This is because the $d_{xy}$ orbitals form non-bonding states without mutual connection via oxygen, while the $d_{yz}$ and $d_{zx}$ orbitals have such connections along the zigzag chain to form the bonding and antibonding states around −2 and 2 eV, respectively. As a result, two electrons occupy the bonding $d_{yz}$ and $d_{zx}$ states, and the remaining electron fills the non-bonding $d_{xy}$ states just below $E_F$. Thus, the Fermi surfaces of β-ReO$_2$ originate from the nearly half-filled $d_{xy}$ states.

Figure 7b shows the calculated Fermi surfaces visualized with the FermiSurfer application.[57] There are three large and one small Fermi surfaces. The large electron Fermi surface (#101 electron) has an ellipsoidal shape surrounding the $\Gamma$ point with a horn extending to the $U$ point at $k_z = \pi$. The two hole Fermi surfaces (#97 hole and #99 hole) are located surrounding the $R$–$T$ line at the zone boundary. And the tiny electron pocket (#103 electron) exists around the $U$ point. Taking all these Fermi surfaces into account, we calculated the Hall coefficient based on the Boltzmann theory with the constant relaxation approximation[58–60] to be $R_\mathrm{H} = 5\times10^{-4}$ cm$^3$ C$^{-1}$, which is in reasonable agreement with the experimental value of $8.1\times10^{-4}$ cm$^3$ C$^{-1}$ at 2 K. The large electron and hole Fermi surfaces must have a dominant contribution to bulk properties such as magnetic susceptibility and heat capacity because of the large DOS at $E_\mathrm{F}$. On the other hand, the Fermi surface responsible for the observed quantum oscillations must be the small electron pocket #103. In fact, the calculated extremal cross section perpendicular to the $k_x$ axis and the effective mass of #103 electron are 73 T and $0.26m_0$, respectively, which are in good agreement with the experimental values of 51 T and $0.23m_0$.

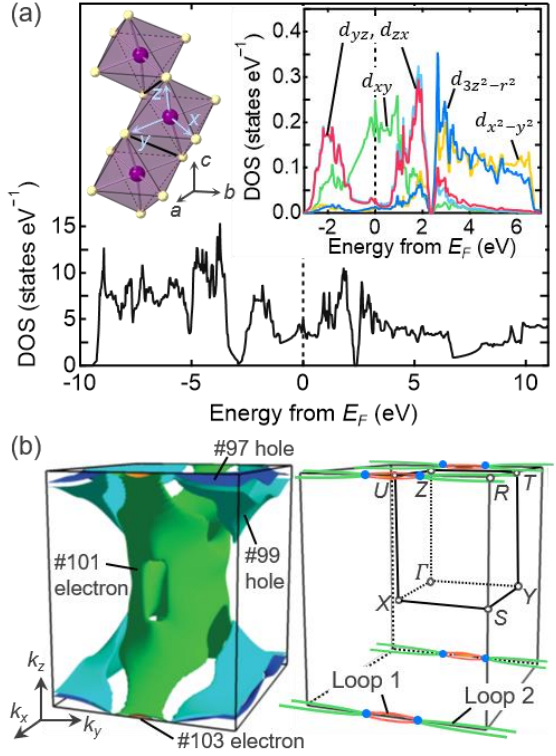

**Fig. 7.** (Color online) (a) Density of states (DOS) (main panel) of β-ReO$_2$, and the partial DOS of Re 5$d$ orbitals (right inset). The local axes ($x$, $y$, $z$) for the 5d orbitals are presented in the left inset. (b) Calculated Fermi surfaces of β-ReO$_2$ consisting of four Fermi surfaces (#97 hole, #99 hole, #101 electron, #103 electron). At the edge of the Brillouin zone along the $U$–$R$ line, there is a Dirac loop chain made of loop 1 (red circle) surrounding the $U$ point in the $k_z = \pi$ plane and loop 2 (green circle) surrounding the $R$ point in the $k_x = \pi$ plane. The small electron pocket (#103 electron) is located inside of loop 1.

The most important feature of the electronic structure of β-ReO$_2$ is the presence of DLC. According to the previous calculations,[28] there are two types of Dirac loops in the $k$-space that are protected by the $n$- and $b$-glide symmetries as well as time-reversal symmetry: loop 1 centered at the $U$ point on the $k_z = \pi$ plane and loop 2 centered at the $R$ point on the $k_x = \pi$ plane; they should be orthogonal to each other, because they are generated by the $n$- and $b$-glide planes, respectively. Moreover, they are connected at a point on the $U$–$R$ line to form a DLC along the $k_y$ direction. On the other hand, the remaining $a$-glide plane does not cause a Dirac loop but a Dirac point on the $T$–$Y$ line. Our calculations confirmed these features, and, moreover, revealed their actual shape, which is quite thin as depicted in Fig. 7b: loop 1 crosses the $U$–$Z$ line at $k_x = 0.984\pi$, loop 2 crosses the $R$–$S$ line at $k_z = 0.986\pi$, and the intersection point of the two loops are located at ($\pi$, $0.26\pi$, $\pi$). It is important to note that the Dirac point at the intersection lies only 15 meV below $E_\mathrm{F}$, which is much smaller than those in other Dirac semimetals: 230 meV for Cd$_3$As$_2$[21] and 30 meV for high-quality Na$_3$Bi.[19] Therefore, one expects that the DLC-derived quasiparticles have a significant influence on the transport properties of β-ReO$_2$. Especially, loop 1 may be related to the observed XMR because it is located around the #103 electron Fermi surface.

## 4. Discussion

We have found that a high-quality single crystal of β-ReO$_2$ exhibits an XMR of 22,000% at 10 T and 2 K in spite that it seems to be a conventional Pauli paramagnetic metal with a high carrier density of $1\times10^{22}$ cm$^{-3}$ and considerable mass enhancements in the Sommerfeld coefficient and magnetic susceptibility. Moreover, we observed dHvA oscillations with frequencies of 51 and 79 T and effective masses of $0.23m_0$ and $0.40m_0$, respectively, which evidences the presence of small Fermi surfaces. Our band structure calculations revealed one large electron and two hole Fermi surfaces, which may dominate the fundamental properties, and one small electron Fermi surface (#103 electron) which is responsible for the dHvA oscillations. Furthermore, they clarified the existence of a thin DLC made of two Dirac loops along the $k_y$ direction, with loop 1 surrounding the #103 electron pocket just below $E_\mathrm{F}$.

XMR has been commonly observed in topological semimetals called Dirac and Wyle semimetals. For example, the following compounds were studied: WTe$_2$ ($1.7\times10^5$% at 2 K and 9 T),[44] LaSb ($1.2\times10^5 \sim 9.0\times10^5$ % at 2 K and 9 T),[61] PtSn$_4$ ($2.6\times10^5$% at 1.8 K and 10 T),[62] Cd$_3$Ad$_2$ ($3.5\times10^3 \sim 1.4\times10^5$ % at 5 K and 9 T),[21] NbP ($8.5\times10^5$% at 1.85 K and 9 T).[22] Table I compares β-ReO$_2$ with the typical Dirac semimetal Cd$_3$As$_2$ and the Weyl semimetal NbP, which show one-order larger XMR than β-ReO$_2$. The carrier densities of the two compounds are very low in the order of $10^{18}$ cm$^{-3}$, while the mobilities are as high as $10^6$ cm$^2$ V$^{-1}$ s$^{-1}$, which results in the large conductivities at low temperature. This is obviously due to the linear band dispersions.

The carrier density of β-ReO$_2$ is $7.7\times10^{21}$ cm$^{-3}$, much larger than those of NbP and Cd$_3$As$_2$, and those of WTe$_2$,

Table I. Comparison of transport parameters among XMR compounds with various topological states.

| Material | Topological state | $n$ (cm$^{-3}$) | $\rho_{xx}$ (nΩ cm) | $\mu$ (cm$^2$ V$^{-1}$ s$^{-1}$) | MR (%) | Ref. |
|---|---|---|---|---|---|---|
| Cd$_3$As$_2$ | Dirac semimetal | 7.4×10$^{18}$ | 21 (5 K) | 8.7×10$^6$ | 1.3×10$^5$ (5 K, 9 T) | 21) |
| NbP | Weyl semimetal | 1.5×10$^{18}$ | 630 (2 K) | 5×10$^6$ | 8.5×10$^5$ (1.85 K, 9 T) | 22) |
| β-ReO$_2$ | Hourglass Dirac loop chain metal | 7.7×10$^{21}$ | 206 (2 K) | 3.8×10$^3$ | 2.2×10$^4$ (2 K, 10 T) | This work |

LaSb, PtBi$_2$ (~1×10$^{20}$ cm$^{-3}$) and Na$_3$Bi (~1×10$^{17}$ cm$^{-3}$).[19,21,22,44,56,61] This is apparently due to the presence of additional large Fermi surfaces in β-ReO$_2$. Since the electrical conductivity is given by $\sigma = ne^2\tau/m^*$, where $\tau$ is the scattering lifetime, it is dominated by a large Fermi surface with high carrier density when scattering of carrier is large, while can be influenced by a small Fermi surface with high mobility even with low carrier density. In β-ReO$_2$, the pronounced decrease in resistivity below 100 K is likely due to a crossover in carriers originating from the large to small Fermi surfaces upon cooling. The XMR observed at low temperatures is obviously the characteristic of the light quasiparticles coming from the small Fermi surface. The average mobility at 2 K was estimated to be $\mu(2\,\text{K}) = R_\text{H}(2\,\text{K})/\rho_{xx}(2\,\text{K}) = 3.75\times10^3$ cm$^2$ V$^{-1}$ s$^{-1}$, but the actual mobility of the light quasiparticles must be much larger. Therefore, β-ReO$_2$ is a unique system having both large and small Fermi surfaces, the latter of which is responsible for the XMR.

Here we discuss the origin of the XMR in β-ReO$_2$. Two main mechanisms thus far proposed for the origin of XMR in topological semimetals are the conventional hole–electron compensation mechanism and the violation of topological protection by magnetic fields.[21,44] It is pointed out that the former plays a major role in classical low-carrier semimetals such as Bi and graphite,[63–67] and more recently in WTe$_2$.[44] In this mechanism, the MR is maximized when the holes and electrons are even in the carrier density and decreases rapidly as the carrier densities are unbalanced.[44] This mechanism is not applicable to β-ReO$_2$, because it is not a low-carrier compensated semimetal. Furthermore, the MR in β-ReO$_2$ does not show $B^2$ dependence as expected in this scenario; MR is expressed as $\mu_e\mu_h B^2$ for a compensated semimetal, where $\mu_e$ and $\mu_h$ are the mobilities of electrons and holes.

Another mechanism, the violation of topological protection, has been discussed to explain the XMR in many topological semimetals.[19,21] Quasiparticles with linear dispersion near $E_\text{F}$ can cause large conductivity, because the backscattering of them is significantly suppressed. When a magnetic field is applied, this suppression of scattering is lost, so that the conductivity is reduced. Since the original conductivity of quasiparticles with linear dispersion is very large owing to the high mobility, the MR can be quite large. In β-ReO$_2$, Dirac loop 1 is located near $E_\text{F}$, and its quasiparticles must contribute to the conductivity. Thus, similarly as in other topological semimetals, the violation of topological protection by magnetic field should manifest itself in the XMR of β-ReO$_2$. On the other hand, a $B$-linear MR was theoretically predicted for Dirac systems,[43] and was experimentally observed in several candidates.[21,56,68,69] The observed $B^{1.3}$ dependence in β-ReO$_2$ at low fields may be a signature of Dirac points, but it would be interesting if this specific field dependence is somehow related to the DLC. Further experimental and theoretical studies are required to identify the origin of XMR and the role of the DLC in β-ReO$_2$.

Finally, we address the significant difference in resistivity between the previous and present crystals. The RRR of the previous crystal is two orders of magnitude smaller than those of the present crystals and show no such a marked decrease below 100 K.[31] We suspect that the origin of the difference is the energy shift of $E_\text{F}$ caused by nonstoichiometry. In simple transition metal oxides, oxygen deficiencies are often introduced, which cause a change in the valence of metal ions. Oxygen deficiency results in doping electrons with the $E_\text{F}$ shifting to higher energy. In the present crystal, oxygen deficiency is expected to be small because the crystal was grown under relatively oxidizing conditions, as evidenced by the coprecipitation of ReO$_3$ crystals deposited on the crystal surface of β-ReO$_2$. Previous synthesis conditions were possibly more reductive than in this study. When the $E_\text{F}$ shifts to higher energy and moves away from the DLC, the influence of DLC on the transport properties is reduced, eliminating the large conductivity at low temperatures and XMR. This comparison, on the contrary, suggests that the DLC does indeed play an important role in the transport properties of stoichiometric β-ReO$_2$.

## 5. Conclusion

We grew high-quality single crystals of β-ReO$_2$ with a nonsymmorphic crystal structure and observed an XMR of 22,000% at 10 T and 2 K. The carrier density obtained from Hall measurements is about 1×10$^{22}$ cm$^{-3}$, which is much higher than those of typical semimetals that exhibit XMR. The band structure calculations revealed the existence of large Fermi surfaces, indicating that the origin of XMR is not ascribe to the conventional hole–electron compensation. On the other hand, a small Fermi surface was observed in quantum oscillations and supported by band structure calculations. Interestingly, there is a Dirac loop with the hourglass dispersion protected by the glide symmetry, which is located around the small Fermi surface near $E_F$. It is suggested that high mobility carriers originating from the Dirac loop and the violation of topological protection are responsible for the XMR of β-$ReO_2$.

In future studies, novel properties derived from the DLC would be of great interest. For example, anisotropies in the MR and Hall coefficient may appear depending on the magnetic field direction to the Dirac loops, which is not expected for the isotropic Dirac point. Moreover, a drumhead-type surface state predicated for a DLC metal[28] is to be investigated, which may lead to a large enhancement of Friedel oscillation, surface superconductivity and surface ferromagnetism owing to the large DOS. Surface sensitive measurements such as scanning tunneling microscope/spectroscopy and photoemission spectroscopy are planned. From the material point of view, β-$ReO_2$ is much more stable than other compounds that exhibit XMR, which is advantageous in both basic researches and applications. Our findings may trigger further search for topological materials based on oxides.

**Acknowledgements** The authors are grateful to I. Tateishi and M. Ogata for insightful discussion. This work was partly supported by Japan Society for the Promotion of Science (JSPS) KAKENHI Grant Numbers JP18H04308 (J-Physics), JP19H04688, JP20H05150 (Quantum Liquid Crystals) and JP20H01858.

*Note added in proof.*—It was reported very recently that $ReO_3$ showed an XMR of ~10,000% at 9 T and 10 K, although it is an "ordinary" nonmagnetic metal without topological characters[70].

*dhirai@issp.u-tokyo.ac.jp